\documentclass[journal]{IEEEtran}
\usepackage{cite}
\usepackage{amsmath,amssymb,amsfonts}
\usepackage{algorithmic}
\usepackage{graphicx}
\usepackage{algorithm}
\usepackage{textcomp}
\usepackage{multirow}

\usepackage{xcolor}
\usepackage{color}

\ifCLASSINFOpdf
\else
\fi
\begin{document}
\title{MS Lesion Segmentation: Revisiting Weighting Mechanisms for Federated Learning}


\author{Dongnan~Liu,
        Mariano~Cabezas,
        Dongang~Wang,
        Zihao~Tang,
        Lei~Bai,
        Geng~Zhan,
        Yuling~Luo,
        Kain~Kyle,
        Linda~Ly,
        James~Yu,
        Chun-Chien Shieh,
        Aria~Nguyen,
        Ettikan Kandasamy Karuppiah,
        Ryan~Sullivan,
        Fernando~Calamante,
        Michael~Barnett,
        Wanli~Ouyang,
        Weidong~Cai,
        and~Chenyu~Wang
\thanks{D. Liu, and Z. Tang are with the School of Computer Science, The University of Sydney, NSW
2008, Australia, and also with Brain and Mind Centre, The University of Sydney, Sydney 2050, Australia.}
\thanks{M. Cabezas is with Brain and Mind Centre, The University of Sydney, Sydney 2050, Australia.}
\thanks{D. Wang, G. Zhan, Y. Luo, K. Kain, L. Ly, J. Yu, C-C. Shieh, A. Nguyen, M. Barnett, and C. Wang are with Brain and Mind Centre, The University of Sydney, Sydney 2050, Australia, and also with Sydney Neuroimaging Analysis Centre, Sydney 2050, Australia.}
\thanks{L. Bai is with the School of Electrical and Information Engineering, The University of Sydney, NSW 2008, Australia, and also with Brain and Mind Centre, The University of Sydney, Sydney 2050, Australia.}
\thanks{E. K. Karuppiah, is with NVIDIA Corporation, Singapore}
\thanks{R. Sullivan, is with the School of Biomedical Engineering, The University of Sydney, Sydney 2050, Australia}
\thanks{F. Calamante, is with the School of Biomedical Engineering, The University of Sydney, Sydney 2050, Australia, and also with Sydney Imaging, The University of Sydney, Sydney 2050, Australia}
\thanks{O. Wanli is with the School of Electrical and Information Engineering, The University of Sydney, Sydney, NSW 2008, Australia.}
\thanks{W. Cai is with the School of Computer Science, The University of Sydney, Sydney, NSW 2008, Australia.}}

\maketitle

\begin{abstract}
Federated learning (FL) has been widely employed for medical image analysis to facilitate multi-client collaborative learning without sharing raw data. Despite great success, FL’s performance is limited for multiple sclerosis (MS) lesion segmentation tasks, due to variance in lesion characteristics imparted by different scanners and acquisition parameters. In this work, we propose the first FL MS lesion segmentation framework via two effective re-weighting mechanisms. Specifically, a learnable weight is assigned to each local node during the aggregation process, based on its segmentation performance. In addition, the segmentation loss function in each client is also re-weighted according to the lesion volume for the data during training. Comparison experiments on two FL MS segmentation scenarios using public and clinical datasets have demonstrated the effectiveness of the proposed method by outperforming other FL methods significantly. Furthermore, the segmentation performance of FL incorporating our proposed aggregation mechanism can exceed centralised training with all the raw data. The extensive evaluation also indicated the superiority of our method when estimating brain volume differences estimation after lesion inpainting.
\end{abstract}


\IEEEpeerreviewmaketitle

\section{Introduction}
\label{sec:introduction}

\IEEEPARstart{M}{ultiple} sclerosis (MS) is a chronic inflammatory and degenerative disease of the central nervous system, characterized by the appearance of focal lesions in the white and gray matter that topographically correlate with an individual patient’s neurological symptoms and signs. Globally there are an estimated $2.3$ million people with MS and, after trauma, the disease constitutes the most common cause of neurological disability in young adults~\cite{coles2008alemtuzumab}. Lesion characteristics, such as number and volume, are principal imaging metrics for both MS clinical trials and monitoring of the disease in clinical practice~\cite{pontillo2021radiomics}. To this end, automatic and accurate MS lesion segmentation in Magnetic Resonance (MR) imaging can critically enhance both MS research and patient management~\cite{zijdenbos2002automatic,elliott2013temporally,brosch2016deep,llado2012segmentation}.

With recent advances on AI-enhanced computer vision techniques, deep learning-based methods have been widely used for lesion segmentation and can achieve results close to human experts. Despite this, there remain significant challenges in the current methods~\cite{ma2022multiple,danelakis2018survey}. In particular, MR images from different scanners present different data distributions, which incur a performance drop when validating off-the-shelf models trained at a single client with images from another. On the other hand, data sharing between multiple clients is not always possible due to privacy, legal and ethical concerns.

\begin{figure}[t]
\centering
\includegraphics[width=0.49\textwidth]{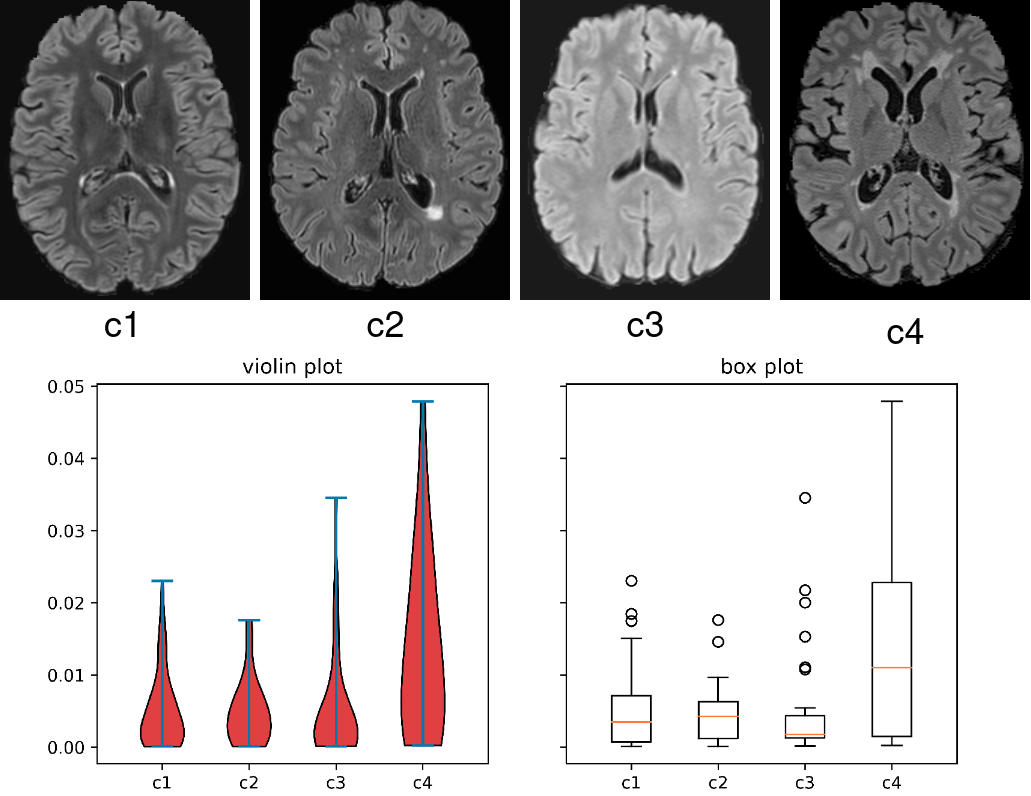} 
\caption{Evidence of the variance on appearance and lesion volume in multi-client studies. The top images are examples of 2D slices from each client in the study. The bottom graphs are the violin and box plots for the lesion volume to brain volume ratio distributions per client for all the subjects in this FL study.}
\label{fig-intro}
\end{figure}

To address this dilemma, federated learning (FL) techniques where training is decentralized were proposed~\cite{mcmahan2017communication,li2021fedbn}. At the beginning of the FL process, each participating client is firstly assigned an initialized model. Next, these models are trained using the local data in each client. After several training iterations, each client is required to share their private model weights with a central server, which aggregates these local weights and distributes them back to each client. Initialized by the updated weights from the server, the model in each client continues their local training for another round of FL process. By enriching the knowledge learned in each local model without sharing the raw data, FL methods have also been widely employed for multi-client medical image analysis~\cite{li2020multi,guo2021multi,liu2021feddg,shen2021multi}. Note that throughout the paper, we use the notion `client' to represent the data in each distinct scanner or clinical center.

However, current FL methods are suboptimal for multi-client MS lesion segmentation. First, during aggregation, the central server averages the model parameters from all the local clients, assuming each local model has the same importance and performance. For MS lesion segmentation, the datasets from multiple clients, their data distribution and the lesion morphology and signal characteristics can vary greatly~\cite{kamnitsas2017unsupervised,ackaouy2020unsupervised}, which can lead to divergence of the private local models, thereby conferring distinct segmentation characteristics when they are aggregated in the central server. By fusing a model with inferior segmentation performance to others with superior ability, the segmentation performance for the entire updated model may be compromised. Second, differences in the clinical distribution of patients can impact lesion burden, size and morphology at a client level, generating significant inter-site variance in multi-client studies, as shown in Fig.~\ref{fig-intro}. As explored in~\cite{nichyporuk2021optimizing,shirokikh2020universal}, a model trained on a dataset with smaller lesions will usually present a lower performance due to the lack of lesion samples for training. However, the task loss functions in each client are optimized with the same importance in previous FL methods~\cite{mcmahan2017communication,li2020multi,li2021fedbn}, which would induce the inferior performance of the center model on the clients with smaller lesion sizes, and further influence the overall FL segmentation accuracy.

To solve the aforementioned issues, we propose a Federated MS lesion segmentation framework based on two dynamic Re-Weighting mechanisms (FedMSRW). During the model aggregation process, the model parameters from each client are assigned a weight based on their segmentation abilities during local training, including the segmentation performance and confidence. Models with higher ability are assigned a higher weight and vice versa. To solve the lesion volume imbalance across different clients, we propose to re-weight the task loss function in each client based on the average case-wise lesion volume ratio, i.e., the ratio of lesion volume to the brain volume, of the training data for that client. Motivated by~\cite{shirokikh2020universal}, where more attention should be paid to smaller lesion objects during model training, the weights for the overall loss functions in clients with a smaller lesion volume are enlarged, and vice versa.

The major contributions of this work are summarized as follows:

\begin{itemize}
    \item To the best of our knowledge, this work is the first application of privacy-preserving FL methods to the task of MS lesion segmentation and, in particular, to multi-client MS datasets.
    \item We propose uncertainty-aware re-weighting mechanisms during the central model aggregation process to prevent the negative influence of the inferior local models.
    \item We further propose to re-weight the segmentation loss functions in each local client/center based on its local lesion volume ratio, addressing the impact of client-specific lesion variance in the multi-client MS datasets.
    \item We have conducted extensive experiments in two FL MS lesion segmentation scenarios using both public and real-world clinical MS datasets. Our FedMSRW method outperforms typical FL methods significantly.
\end{itemize}

\begin{figure*}[t]
\centering
\includegraphics[width=0.95\textwidth]{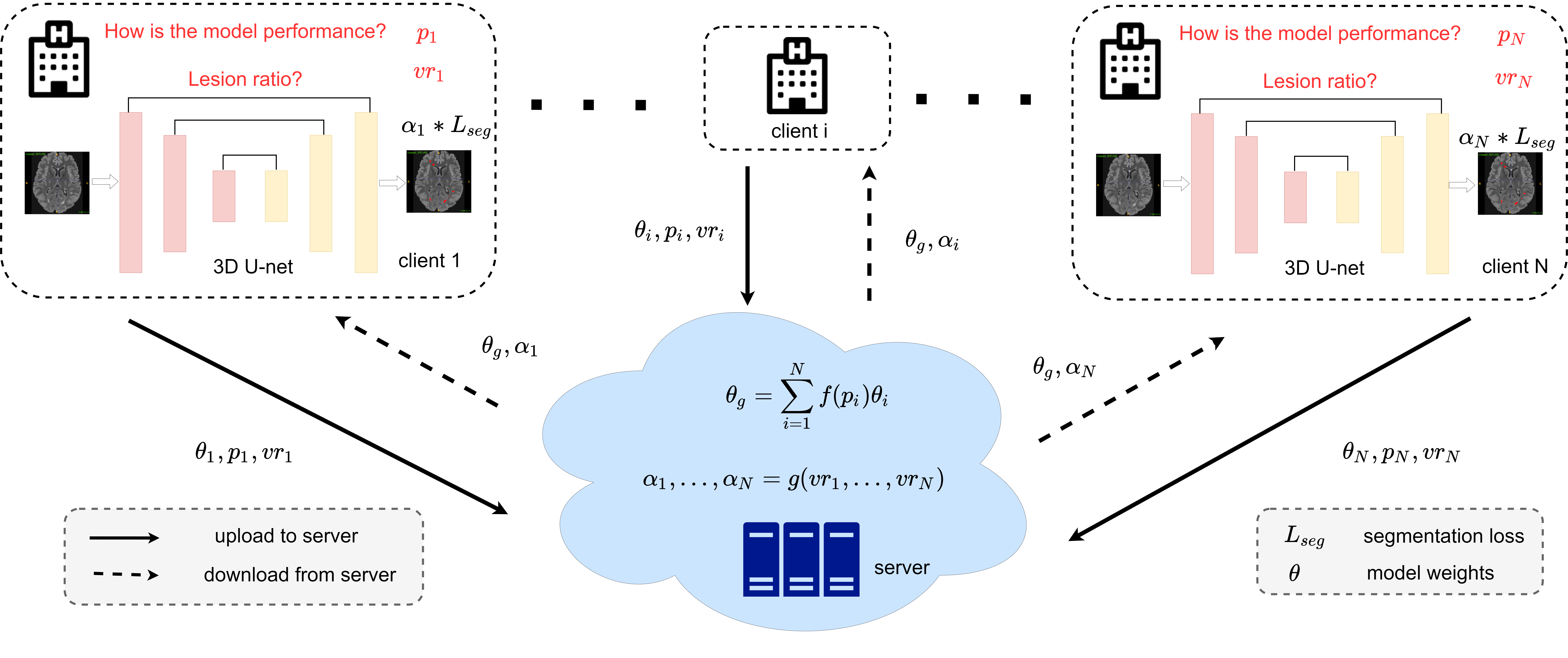} 
\caption{Detailed framework of our FedMSRW method. The $f(.)$ for calculating the weighting factors during model aggregation can be referred to Equation~\ref{equ-rw-agg}. The details of $g(.)$ for the segmentation task re-weighting are in Equation~\ref{equ-dice-rw}.}
\label{fig-method}
\end{figure*}

\section{Related Work}

\subsection{MS Lesion Segmentation}

In classical MS lesion segmentation methods, the brain tissues (e.g., WM, GM, CSF) are firstly segmented from the raw MR images via statistical methods, e.g., the Expectation-Maximization (EM) algorithm~\cite{catanese2015automatic,beaumont2016automatic} or Gaussian Mixture Modeling~\cite{doyle2016automatic,knight2016ms}. Then, lesions are detected as outliers based on the tissue masks~\cite{van2001automated,catanese2015automatic,beaumont2016automatic,doyle2016automatic,knight2016ms}. With the advent of deep learning-based medical image computing~\cite{shin2016deep,shen2017deep}, deep learning models that learn representative features via convolutional modules have been widely employed for automatic MS lesion segmentation, achieving competitive performance~\cite{ghafoorian2017deep,zhang2018ms,nair2020exploring,mckinley2020automatic,isensee2021nnu,ma2022multiple,brosch2016deep,valverde2017improving,mehta2021propagating}. 
 
In clinical practice, the data distribution of brain MRI varies across MRI scanners due to variance in image geometry, resolution, tissue intensity and contrast conferred by differences in hardware (scanner and coil) and acquisition protocols~\cite{kamnitsas2017unsupervised,ackaouy2020unsupervised,valverde2019one,dewey2019deepharmony}. These domain differences limit the performance of supervised learning methods when applied to images from new scanners~\cite{ma2022multiple,kamnitsas2017unsupervised,ackaouy2020unsupervised}. Such phenomenon is referred to as the domain shift issue, which exists in various medical image analyses applications for multiple datasets from different resources (e.g, modalities, sites)~\cite{chen2020unsupervised,liu2020pdam,xu2021cross}. Recently, cross-domain MS lesion segmentation methods have been further explored to enhance the models’ generalization ability. In particular, the domain differences are alleviated by inducing the model to generate scanner-invariant features~\cite{kamnitsas2017unsupervised,ackaouy2020unsupervised}, learning from synthetic images that follow the distribution of the target scanners~\cite{palladino2020unsupervised}, and cross-scanner data harmonization~\cite{dewey2019deepharmony}. A crucial prerequisite of these methods is that all the data from multiple scanners should be fed into the framework simultaneously. However, sharing clinical data across sites invokes privacy issues, which limit the practical applications of these methods in large collaborative studies~\cite{li2020multi,guo2021multi}. Recently, cross-domain MS lesion segmentation methods have been further explored to enhance the models' generalization ability. Particularly, the domain differences are alleviated by inducing the model to generate scanner-invariant features~\cite{kamnitsas2017unsupervised,ackaouy2020unsupervised}, learning from synthetic images that follow the distribution of the target scanners~\cite{palladino2020unsupervised}, and cross-scanner data harmonization~\cite{dewey2019deepharmony}. A crucial prerequisite of these methods is that all the data from multiple scanners should be fed into the framework simultaneously. However, sharing clinical data across sites involves privacy issues, which limits the practical applications of these methods on large collaborative studies~\cite{li2020multi,guo2021multi}.

\subsection{Federated Learning}

Federated learning (FL) provides a decentralized solution for multi-client collaborative learning without raw data sharing. To ensure privacy preservation, a central server is established to collect, from each local client, their model weights, gradients, and features. Next, an aggregation process is conducted to update the central model, which is then broadcasted back to each local client. Originally, FedAvg~\cite{mcmahan2017communication} performed the aggregation through averaging the weights from all local clients. This approach was later extended~\cite{mcmahan2017communication} to propose FedProx~\cite{li2020federated}, with regularization to stabilize the models’ performance.

FL methods have also been utilized for multi-client medical image analysis. In~\cite{li2020multi} and~\cite{guo2021multi}, each local model is incorporated with an adversarial domain discriminator to alleviate the inter-client distribution bias. However, the intermediate features in each local client are required to be shared across clients. Despite these privacy-preserving strategies, distributing features still incur the risk of data leakage. To solve this problem, FedBN~\cite{li2021fedbn} has been proposed for domain adaptive FL by only processing the parameters outside the batch normalization layers of each local model. In addition, FedDG~\cite{liu2021feddg} enhances the generalization ability of the FL framework on the unseen datasets via Fourier transform-based image synthesis and episodic learning strategies. Furthermore,~\cite{gong2021ensemble} proposed a distillation-based FL method without sharing the model parameters, which further enhances data safety. Although these methods are effective in many medical imaging scenarios, they have not considered the weighting strategies for the global aggregation and local training, which is crucial for FL MS segmentation. The method that is most related to ours is~\cite{shen2021multi}, which re-weights each local model’s training based on the loss value changes. However, its dynamic weighting strategy is sensitive to hyperparameter selections, which lacks robustness. Rather, our proposed re-weighting mechanisms at the global and local levels are effective and simple, without auxiliary hyperparameters.

\section{Methods}

In this section, we first present the overview of our proposed FedMSRW method. Next, the re-weighting mechanisms during the central aggregation and local training are respectively illustrated in detail. Finally, we introduce the training and inference details of the FedMSRW method.

\subsection{Overview}


We denote $D_{i}=\{(X_{i}, Y_{i}\}_{i = 1, 2,...,N}$ as the set of MS lesion segmentation datasets from $N$ different clients, where $X$ and $Y$ represent the MR images and the corresponding lesion annotations. In the $i~th$ client, the local model $M_{i}$ with the parameters $\theta_{i}$ is optimized via:

\begin{equation}
\begin{aligned}
L_{i} = \min\limits_{\theta_{i}} L_{dice}\big(M_{i}(X_{i}), Y_{i}\big),
\label{equ-dice}
\end{aligned}
\end{equation}
where $L_{dice}$ is the soft Dice loss function for probabilistic binary segmentations~\cite{milletari2016v}:

\begin{equation}
\begin{aligned}
L_{dice} = 1 - \frac{2 \sum M_{i}(X_{i}) Y_{i}}{ \sum M_{i}(X_{i})^2 + \sum Y_{i}^2}.
\label{equ-dice-details}
\end{aligned}
\end{equation}

Due to the data distribution differences in multi-client MR images, we establish our proposed FedMSRW on FedBN~\cite{li2021fedbn}, which tackles the domain bias issues in FL processes that only require sharing of the model parameters. Based on the assumption that the parameters of the normalization layers in deep learning models represent the domain-specific information~\cite{chang2019domain,huang2018multimodal}, FedBN prevents the central model from domain shift by aggregating the parameters in the convolutional layers, while ignoring those in the batch normalization layers. Specifically, each $\theta_{i}$ can be represented as: $\theta_{i} = \{\theta_{i}^{bn},  \theta_{i}^{r} \}$, where $\theta_{i}^{bn}$ are the parameters for all the batch normalization layers, and $\theta_{i}^{r}$ are those for the rest layers. After collecting the local weights, the central server aggregates model through:
\begin{equation}
\begin{aligned}
\hat{\theta}^{r} = \frac{1}{N}\sum_{i}^{N}\theta_{i}^{r}.
\label{equ-fedbn}
\end{aligned}
\end{equation}
By receiving the updated weights from the server, each $M_{i}$ is then initialized as:  $\hat{\theta}_{i} = \{\theta_{i}^{bn},  \hat{\theta}^{r} \}$, for the next round of local segmentation training. The detailed framework is shown in Fig.~\ref{fig-method}.


\subsection{Central Aggregation Re-weighting based on the Models' Segmentation}


Due to distinct, client-specific characteristics of both the MRI data and the MS lesions, the difficulty of lesion segmentation tasks differs across clients. To this end, the segmentation ability for the various $M_{i}$ is different after each round of local training. According to Equation~\ref{equ-fedbn}, both the low-performance and high-performance models are assigned equal importance during the aggregation process at the central server. This is suboptimal since the local models with inferior segmentation ability influence the updated model from the server and further limit collaborative knowledge learning in FL. A trivial solution to this problem is to adjust the number of training samples for each client, as indicated in~\cite{mcmahan2017communication}. However, there is no simple, non-biased sample selection mechanism to alleviate the negative effects of the models with inferior performance. Additionally, selecting auxiliary hyperparameters manually in FL would limit the model’s robustness.

To this end, we propose an aggregation re-weighting mechanism based on the segmentation performance of each $M_{i}$ during the training process in the local clients. For each training iteration in client $i$, we define the input data and corresponding labels as $x$ and $y$, respectively. The segmentation ability for probabilistic lesion segmentation $M_{i}$ is measured as:

\begin{equation}
\begin{aligned}
P_{i} = \frac{\sum M_{i}(x) * y}{\sum y}  * (1- L_{dice}\big(M_{i}(x), y\big)).
\label{equ-rw-agg}
\end{aligned}
\end{equation}
As indicated in Equation~\ref{equ-rw-agg}, the first item represents the models' confidence in the predicted lesion segmentation. Since the MS lesion region of interest occupies only a tiny fraction (average around $1\%$) of the whole brain volume, the confidence value within the true positive lesion regions better reflects the models' lesion prediction certainty relative to traditional methods that measure the models' confidence based on the entropy of the whole prediction map. Additionally, the $(1- L_{dice}\big(M_{i}(x), y\big))$ in Equation~\ref{equ-rw-agg} on the model's segmentation performance is further included. Finally, the average $P_{i}$ for all the local training iterations is able to indicate the segmentation ability for the $M_{i}$. Considering $P_{i}$, the central aggregation process in Equation~\ref{equ-fedbn} is re-formulated as:

\begin{equation}
\begin{aligned}
\hat{\theta}^{r}_{rw} = \frac{1}{\sum P_{i}}\sum_{i}^{N}\theta_{i}^{r} * P_{i}.
\label{equ-fedbn-rw-agg}
\end{aligned}
\end{equation}

\begin{table*}[!htb]
\centering
\caption{The comparison experiments between our proposed method and others on the first FL MS lesion segmentation scenario, using MICCAI MSSEG16 dataset.}
\resizebox{0.99\linewidth}{!}{%
\begin{tabular}{c| c c c c | c c c c | c c c c | c c c c }
\hline
\multirow{2}{*}{} &  \multicolumn{4}{c|}{$C-Dice\uparrow$}  & \multicolumn{4}{c|}{$V-Dice\uparrow$}  &  \multicolumn{4}{c|}{$V-TPR\uparrow$}  &  \multicolumn{4}{c}{$V-FPR\downarrow$} \\
\cline{2-17}
 & $C1$ & $C2$  & $C3$ & $avg$ & $C1$  & $C2$ & $C3$ & $avg$  & $C1$ & $C2$ & $C3$  & $avg$  & $C1$ & $C2$  & $C3$ & $avg$ \\
  \hline
Single & $ 61.98  $ &  $ 41.00  $ &  $71.32 $  &  $ 58.10$ & $ 76.88  $ &  $22.17  $ &  $ 77.20 $  &  $58.75 $ &  $68.94$ &  $ 12.97  $ &  $ 70.54 $  &  $50.82$ &  $ 13.12  $ &  $ 23.68  $  &  $ 14.76$  &  $ 17.19 $\\
Central & $ 69.20 $ &  $51.44   $ &  $69.84 $  &  $ 63.49 $ & $ 77.04 $ &  $56.71   $ &  $ 78.09 $  &  $70.62 $ &  $70.63$ &  $ 45.82 $ &  $ 74.01$  &  $ 63.49$ &  $15.27 $ &  $25.60 $  &  $ 17.35$  &  $ 19.41 $\\
 \hline
FedAvg  & $ 42.98   $ &  $45.12   $ &  $ 70.60 $  &  $52.90 $ & $ 54.73  $ &  $ 28.02 $ &  $ 76.74 $  &  $ 53.16 $ &  $ 38.78$ &  $\textbf{58.64}  $ &  $81.17 $  &  $59.53$ &  $ \textbf{7.01}  $ &  $ 81.59  $  &  $ 27.24 $ &  $ 38.61  $ \\

FedProx  & $ 39.11  $ &  $ 48.60  $ &  $70.87 $  &  $52.86  $ & $ 53.28   $ &  $ 43.95  $ &  $ 76.47 $  &  $ 57.90 $ &  $37.49$ &  $ 51.67 $ &  $ 74.84 $  &  $ 54.67  $ &  $ 7.98  $ &  $ 61.76  $  &  $ 21.83  $ &  $ 30.52 $ \\

FedBN  & $ 57.18  $ &  $ 53.42  $ &  $ 68.83 $  &  $59.81  $ & $ 70.39  $ &  $ 47.60  $ &  $ 78.09 $  &  $ 65.36  $ &  $ 62.84$ &  $ 33.84  $ &  $  \textbf{82.01} $  &  $ \textbf{59.56}  $ &  $ 20.00  $ &  $ 19.75 $  &  $ 25.47  $  &  $ 21.74  $ \\
DWA & $ 56.13 $ &  $\textbf{58.13} $ &  $ 68.73 $  &  $61.00 $ & $ 66.83 $ &  $\textbf{54.12}  $ &  $75.43$  &  $65.46 $ &  $ 54.43$ &  $ 39.41$ &  $ 73.18$  &  $ 55.67$ &  $ 13.45 $ &  $ 13.66 $  &  $ 22.18  $  &  $  \textbf{16.43} $ \\
  \hline
Ours & $ \textbf{62.76} $ &  $ 56.76$ &  $ \textbf{71.15}$ &  $ \textbf{63.56}$ & $ \textbf{72.60} $ &  $ 51.46$ &  $ \textbf{78.13}$ &  $ \textbf{67.39}$ &  $ \textbf{64.97}$ &  $ 36.65 $ &  $74.58$ &  $ 58.74$ &  $ 17.75 $ &  $ \textbf{13.66} $ &  $ \textbf{17.98} $ &  $16.46 $ \\
 \hline
\end{tabular}}
\label{table-cmp-msseg}
\end{table*}

\begin{table*}[!htb]
\centering
\caption{The comparison experiments between our proposed method and others on the second FL MS lesion segmentation scenario.}
\resizebox{0.99\linewidth}{!}{%
\begin{tabular}{c| c c c c c| c c c c c| c c c c c| c c c c c}
\hline
\multirow{2}{*}{} &  \multicolumn{5}{c|}{$C-Dice\uparrow$}  & \multicolumn{5}{c|}{$V-Dice\uparrow$}  &  \multicolumn{5}{c|}{$V-TPR\uparrow$}  &  \multicolumn{5}{c}{$V-FPR\downarrow$} \\
\cline{2-21}
 & $C1$ & $C2$  & $C3$ & $C4$ & $avg$ & $C1$  & $C2$ & $C3$ & $C4$ & $avg$  & $C1$ & $C2$ & $C3$ & $C4$ & $avg$  & $C1$ & $C2$  & $C3$ & $C4$ & $avg$ \\
 \hline
 Single & $ 55.20   $ &  $ 45.69   $ &  $ 41.92  $  &  $ 58.22  $ & $ 50.26   $ &  $63.33   $ &  $ 40.28  $  &  $43.86 $ &  $69.35 $ &  $54.21  $ &  $ 58.27$  &  $52.53$ &  $ 52.49  $ &  $ 62.37 $  &  $ 56.41$  &  $ 30.64$  &  $ 67.33$ &  $ 62.33 $ &  $ 21.91  $  &  $45.55 $ \\
 Central & $ 55.33   $ &  $ 57.63   $ &  $ 48.84  $  &  $ 48.83 $ & $ 52.66   $ &  $64.31 $ &  $ 65.99  $  &  $48.00 $ &  $55.47 $ &  $ 58.44  $ &  $ 56.08 $  &  $59.74$ &  $53.89 $ &  $40.74 $  &  $52.62$  &  $ 24.64 $  &  $ 26.29$ &  $ 56.73 $ &  $ 13.12 $  &  $ 30.20 $ \\
  \hline
 FedAvg & $  \textbf{52.70}   $ &  $  \textbf{60.37}   $ &  $38.71  $  &  $36.47  $ &  $ 47.06  $ & $ 57.29   $ &  $67.92   $ &  $ 28.32  $  &  $ 37.58  $ &  $ 47.78 $ &  $ 67.04  $ &  $ 63.22 $  &  $63.18  $ &  $ 23.48   $ &  $ 54.23  $  &  $ 49.98 $ &  $  \textbf{26.63} $  &  $ 81.75 $ &  $ \textbf{5.94}   $ &  $ 41.07  $   \\
 
 FedProx  & $ 47.23   $ &  $54.62   $ &  $ 40.79  $  &  $33.59  $ & $44.06   $ &  $55.63   $ &  $ 61.89  $  &  $ 47.49  $ &  $ 33.42 $ &  $49.60  $ &  $ \textbf{68.83} $  &  $ \textbf{66.96}  $ &  $\textbf{ 63.98 }  $ &  $24.21  $  &  $55.99  $  &  $53.32 $  &  $42.47  $ &  $62.25   $ &  $ 46.10  $  &  $ 51.03 $\\

 FedBN  & $51.43  $ &  $51.64   $ &  $ 47.00  $  &  $48.15  $ & $ 49.55   $ &  $ 60.30   $ &  $57.77  $  &  $ 56.63  $ &  $ 55.17 $ &  $57.47  $ &  $ 62.57 $  &  $63.78  $ &  $ 54.62   $ &  $ 40.15  $  &  $55.28 $  &  $ 41.81 $  &  $ 47.20  $ &  $41.20   $ &  $11.89  $  &  $ 35.52  $\\

  DWA & $ 41.88  $ &  $ 37.63 $ &  $ 40.60 $  &  $46.59 $ & $ 41.68  $ &  $46.23  $ &  $28.03 $  &  $ 43.27  $ &  $ 52.20$ &  $ 42.43 $ &  $ 68.84$  &  $ 65.87 $ &  $ 56.12 $ &  $ 39.95 $  &  $ \textbf{57.70}  $  &  $ 65.19 $  &  $ 82.19 $ &  $ 64.79 $  &  $ 24.73  $  &  $ 59.23$  \\
    \hline
Ours  & $52.42 $ &  $ 58.90 $ &  $ \textbf{50.90}$ &  $ \textbf{52.41}$ & $ \textbf{53.66} $ &  $ \textbf{64.22}$ &  $ \textbf{69.48}$ &  $ \textbf{56.90}$ &  $ \textbf{58.61}$ &  $ \textbf{62.31} $ &  $64.17$ &  $ 66.02$ &  $ 54.32 $ &  $ \textbf{45.14} $ &  $ 57.41 $ &  $ \textbf{35.73}$ &  $26.67$ &  $ \textbf{40.25} $ &  $ 16.46 $ &  $ \textbf{29.78} $\\
 \hline
\end{tabular}}
\label{table-cmp-ims}
\end{table*}

\begin{table}[!htb]
\centering
\caption{Details on the scanners for the datasets used in our experiments. }
\resizebox{0.25\textheight}{!}{%
\begin{tabular}{l|l|l}
\hline
Client  & scanner  & cases \\
\hline
\multicolumn{3}{c}{Scenario~1} \\
\hline
C1  &  Siemens Verio 3T  & 5 \\
\hline
C2  & Siemens Aera 1.5T & 5 \\
\hline
C3  & Philips Ingenia 3T & 5 \\
\hline
\hline
\multicolumn{3}{c}{Scenario~2} \\
\hline
C1  & GE Discovery 3T & 54 \\
\hline
C2  & Philips Ingenia 3T & 21 \\
\hline
C3  & Siemens Skyra 3T & 30 \\
\hline
C4  & Siemens Magnetom 3T & 30 \\
\hline

\end{tabular}}
\label{table-data-details}
\end{table}

\subsection{Local Optimization Re-weighting based on the Lesion Volume}


Another challenge in FL MS lesion segmentation tasks is the heterogeneity of lesion size across different clients. As indicated in~\cite{nichyporuk2021optimizing,shirokikh2020universal}, lesions with smaller sizes should be assigned a larger weight during model training. To this end, we further propose to re-weight the segmentation loss functions in each client defined in Equation~\ref{equ-dice} based on the lesion volume.

For the $k~th$ round of local training in client $i$, we first calculate the average lesion volume ratio $vr_{i}^{K}$ of all the data samples for training. Specifically, the lesion ratio in each training patch is the ratio of the lesion volume to the brain volume. Compared with only counting the voxel number of lesions, the lesion volume ratio can avoid inaccurate estimations when the proportions of the brain volume in some specific training patches are small. Next, the $vr_{i}^{K}$ is accumulated with the average lesion volume ratio from the previous $k-1$ round, denoted as $vr_{i}$. With the increase of $k$, the accumulated $vr_{i}$ can represent the true lesion volume ratio for the data used during the model training process in each client. In the $K+1~th$ round of local training, the segmentation loss in Equation~\ref{equ-dice} is then reformulated as:

\begin{equation}
\begin{aligned}
L_{i}^{rw} = \frac{\sum_{i}^{N} vr_{i}}{N * vr_{i}} * L_{i}.
\label{equ-dice-rw}
\end{aligned}
\end{equation}


\begin{algorithm}
\caption{Algorithm for the proposed FedMSRW method}
\label{algorithm-overall}
\begin{algorithmic}[1]
\REQUIRE ~~\\
$D_{1}, ..., D_{N}$: MS lesion segmentation from $N$ clients. \\
In each $D_{i}$, $M_{i}$ is the CNN model with the parameters $\theta_{i}$. \\
P: the number of FL rounds. \\
Q: the number of local training iterations in each round. \\
\FOR{$p \in [1,P]$}
\FOR{$i \in [1,N]$}
\STATE Initialize the $M_{i}$ with the updated global model.
\STATE Obtain the accumulated lesion volume ratio for $i$.
\STATE Optimize the $M_{i}$ via Eq.~\ref{equ-dice-rw} in Q iterations.
\STATE Obtain the $P_{i}$ which measures the segmentation ability for $M_{i}$ by Eq.~\ref{equ-rw-agg}.
\ENDFOR \\
Aggregate local models in the central servers via Eq.~\ref{equ-fedbn-rw-agg}.
Calculate the re-weighting factors in Eq.~\ref{equ-dice-rw}.
\ENDFOR
\RETURN $\theta_{1},...,\theta_{N}$
\end{algorithmic}
\end{algorithm}

\subsection{Training and Inference Details}

The overall training algorithm of our proposed FedMSRW method is indicated in Algorithm~\ref{algorithm-overall}. In each local client, the lesion segmentation task is trained with a 3D U-Net~\cite{cciccek20163d}. During training, we employ the SGD optimizer with a momentum of $0.9$, a weight decay of $0.0005$, and a learning rate of $0.0002$. After every $800$ training iterations, the local models are sent to the central server for aggregation. During inference, the model in each client is constructed by the central aggregated convolutional weights and the client-private batch normalization weights. Our experiment is implemented with PyTorch~\cite{paszke2017automatic} on 4 RTX 6000 GPU devices.

\section{Experiments}

\begin{figure*}[t]
\centering
\includegraphics[width=0.73\textwidth]{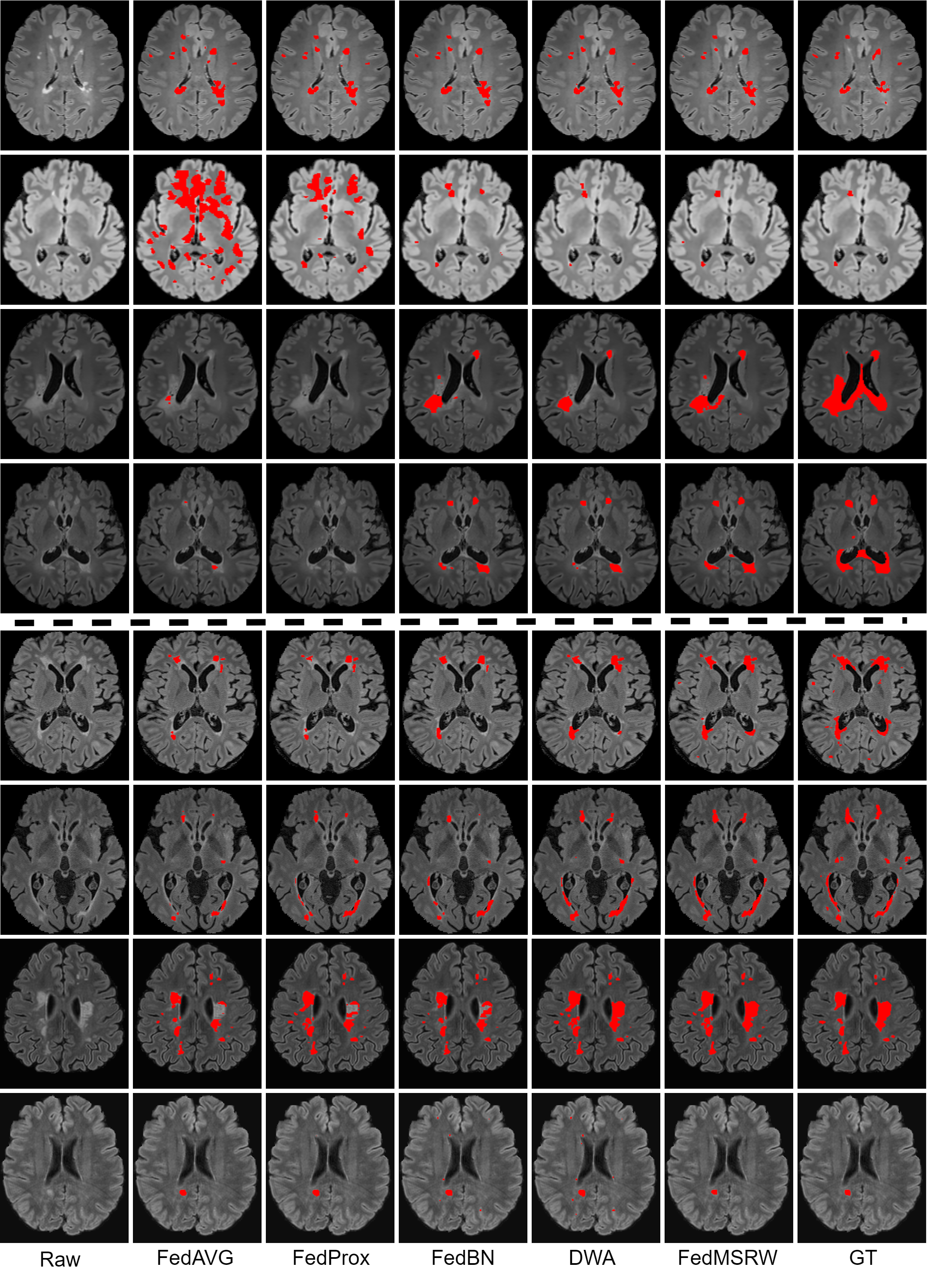} 
\caption{Qualitative results on the comparison FL methods. Lesion masks are overlapped on the original image. The top four rows are the visualization for the Scenario~1, and the bottom four rows are for the Scenario~2.}
\label{fig-cmp}
\end{figure*}

\subsection{Dataset Description}

\subsubsection{Scenario~1}

First, we conducted experiments on the MSSEG16 MS lesion segmentation challenge from MICCAI~\cite{commowick2018objective}. Since the testing images are not publicly available, our experiments were conducted on the training set with 5 cases from 3 different scanners each, as indicated in Table~\ref{table-data-details}. Due to the small number of cases, all experiments were performed in 2-fold cross-validation manners. At each iteration, 3D patches of size $64 \times 64 \times 64$ were randomly cropped from the original FLAIR images, with random flipping and rotation augmentations.

\subsubsection{Scenario~2}

To further indicate the effectiveness of our proposed framework on the FL MS lesion segmentation tasks in a practical clinical scenario, we conducted experiments using in-house and public multi-scanner MS datasets from 4 different scanners. Specifically, the data from C1, C2, and C3 is obtained from de-identified images derived from patients with relapsing and remitting MS, recruited at the Brain and Mind Centre, University of Sydney (Sydney, Australia). All lesion masks were annotated semi-automatically by at least two expert neuroimaging analysts at the Sydney Neuroimaging Analysis Centre (Sydney, Australia). To further increase the diversity of the multi-client MS data, we included a public dataset from a new site~\cite{lesjak2018novel}, in addition to the private data from different scanners. All cases were defaced to protect patient privacy. All experiments under these settings were conducted in a 3-fold cross-validation manner. During training, the $32 \times 32 \times 32$ patches were randomly cropped from the original MRI data, with the augmentations of flipping and rotations.

\subsection{Evaluation Metrics}

To evaluate the segmentation performance of our proposed method, we first employed the case-level and voxel-level Dice coefficient, defined as:

\begin{equation}
\begin{aligned}
Dice = \frac{2TP}{FN + 2TP + FP},
\end{aligned}
\end{equation} 
where TP, FP, and FN indicate the number of true positive, false positive, and false negative voxel predictions, respectively. The case-wise Dice score (C-Dice) was obtained by the average Dice score for all cases, and the voxel-wise Dice score (V-Dice) calculated via the accumulated predictions of all the cases. Additionally, we also evaluated the performance based on the true positive rate (TPR) and false positive rate (FPR) at the voxel level:

\begin{equation}
\begin{aligned}
TPR = \frac{TP}{TP + FN}, FPR = \frac{FP}{TP + FP}.
\end{aligned}
\end{equation}

\subsection{Comparison Experiments}

In this section, we present the experimental results in comparison with typical FL methods, including 1)~\textbf{FedAvg}~\cite{mcmahan2017communication}, a fundamental FL method by central aggregation via averaging of model weights; 2)~\textbf{FedProx}~\cite{li2020federated}, an FL framework introducing an auxiliary regularization mechanism in each client to stabilize learning, 3)~\textbf{FedBN}~\cite{li2021fedbn}, an FL framework which can alleviate the cross-site data distribution bias by ignoring parameters in the normalization layers during aggregation, and 4)~\textbf{DWA}~\cite{shen2021multi}, a dynamic re-weighting mechanism for the central model aggregation process based on the changes of the loss functions in each client. For a fair comparison, we re-implement the DWA on the same FL baseline as our proposed FedMSRW method, i.e., FedBN. We also report the results by training within each local client (\textbf{Single}), and joint training with the raw data from all clients (\textbf{Central}). We maintained the same data split on the N-fold cross-validation for all methods. The experimental results under two FL MS segmentation scenarios are shown in Table~\ref{table-cmp-msseg}, and Table~\ref{table-cmp-ims}, respectively. 

\begin{table*}[!htb]
\centering
\caption{Details of the ablation studies in our experiments. `+ RW-CA' and `+ RW-LT' indicates the FedBN baseline constructed with the proposed central aggregation and local training mechanism, repsectively.}
\resizebox{0.8\linewidth}{!}{%
\begin{tabular}{c|c c c c | c c c c}
\hline
& \multicolumn{4}{c|}{Scenario 1} & \multicolumn{4}{c}{Scenario 2} \\
\cline{2-9}
 & C-Dice $\uparrow$  & V-Dice $\uparrow$  & V-TPR $\uparrow$  & V-FPR $\downarrow$  & C-Dice  $\uparrow$ & V-Dice $\uparrow$ & V-TPR $\uparrow$ & V-FPR  $\downarrow$  \\
\hline
FedBN  & $59.81$ & $65.36$ & $\textbf{59.56}$ & $21.74$  & $49.55$ & $57.47$ & $55.28$ &$35.52$ \\
\hline
+ RW-CA & $60.18$ & $63.21$ & $55.41$ & $24.37$  & $51.96$ & $61.08$ & $57.70$ &$30.35$ \\
\hline
+ RW-LT & $60.95$ & $65.75$ & $57.85$ & $18.31$   & $42.39$ & $45.46$ & $\textbf{58.74}$ &$61.32$ \\
\hline
Ours  & $\textbf{63.56}$ & $\textbf{67.39}$ & $58.74$ & $\textbf{16.46}$   & $\textbf{53.66}$ & $\textbf{62.31}$ & $57.41$ &$\textbf{29.78}$ \\
\hline
\end{tabular}}
\label{table-ablation-all}
\end{table*}

In the first scenario, the MS lesion segmentation experiments were conducted on images from different scanners (from different clinical sites). As shown in Table~\ref{table-cmp-msseg}, the performance of the typical FedAvg and FedProx methods is worse than the models solely trained with the data in each specific client. For the multi-client MS lesion segmentation dataset, the data distributions for each client are distinct, reflecting variance in hardware and image acquisition protocols. This results in domain bias issues when optimizing the aggregated model on each local client. For MS lesion segmentation task, the foreground objects (i.e. lesions) are almost always small and numerous, with a heterogenous spatial distribution. Subsequently, the domain shifts incur inaccurate segmentation performance for the FedAvg and FedProx methods. By preserving the domain-specific batch normalization in each client, FedBN can alleviate the issue and improve the locally trained models. By further prioritizing inter-client label bias and the distinct model performance, our proposed method outperformed the FedBN and DWA, based on the Dice score at both the case and voxel levels. Our proposed FL method also outperformed centralized training (improved average Dice score and competitive voxel-wise Dice score), which requires each client shares their raw data.

In the second scenario, FL methods were conducted on the in-house and public datasets. First, real-world clinical MS datasets from three different scanners were employed. To further increase the diversity of the FL setting, we include an auxiliary public dataset from another new scanner~\cite{lesjak2018novel}. The experimental results are presented in Table~\ref{table-cmp-ims}. We observed a similar phenomenon as the first scenario, namely that cross-client distribution bias in multi-client MS datasets degrades the collaborative performance of the FedAvg and FedProx, while FedBN achieves much better performance by alleviating the domain bias. However, incorporating the DWA with the FedBN baseline has incurred a severe performance drop. The relatively larger dataset used from each client in the second scenario, which exaggerates client-specific differences in data distribution, may explain this observation. Conversely, FedMSRW, which further considers task-specific factors such as cross-client lesion ratios, and distinct local model MS lesion segmentation ability, outperformed the FedBN under all metrics. Fig.~\ref{fig-cmp} illustrates a visual comparison of FedMSRW with other methods, which further indicates the outstanding segmentation performance of our method.

\begin{table*}[!htb]
\centering
\caption{Results on the effectiveness of our proposed FedMSRW under different model designs.}
\resizebox{0.8\linewidth}{!}{%
\begin{tabular}{c|c c c c | c c c c}
\hline
& \multicolumn{4}{c|}{Scenario 1} & \multicolumn{4}{c}{Scenario 2} \\
\cline{2-9}
 & C-Dice $\uparrow$  & V-Dice $\uparrow$  & V-TPR $\uparrow$  & V-FPR $\downarrow$  & C-Dice  $\uparrow$ & V-Dice $\uparrow$ & V-TPR $\uparrow$ & V-FPR  $\downarrow$  \\
\hline
FedBN  & $59.81$ & $65.36$ & $\textbf{59.56}$ & $21.74$  & $49.55$ & $57.47$ & $55.28$ &$35.52$ \\
\hline
Ours-ent & $\textbf{63.70}$ & $65.40$ & $57.98$ & $\textbf{15.83}$ & $41.39$ & $46.91$ & $57.36$ & $59.39$ \\
\hline
Ours-vol & $61.54$ & $63.52$ & $55.34$ & $16.62$ &$46.90$ & $53.03$ & $\textbf{62.26}$ & $52.21$ \\
\hline
Ours  & $63.56$ & $\textbf{67.39}$ & $58.74$ & $16.46$   & $\textbf{53.66}$ & $\textbf{62.31}$ & $57.41$ &$\textbf{29.78}$ \\
\hline
\end{tabular}}
\label{table-select-all}
\end{table*}

\subsection{Ablation Studies}

In contrast to typical FL benchmark tasks, which assume the annotations for each client are in the same distribution space~\cite{li2021fedbn}, the MS lesion segmentation task is confounded by substantial inter-client lesion heterogeneity / distinctions. For specific clients whose MR images generally contain smaller lesions with more noise, it is more challenging for a 3D U-Net to segment lesions accurately. To this end, the effectiveness of the FedBN is still limited by ignoring the bias of labelling space on MS lesion segmentation tasks. To solve this problem, we propose a re-evaluation of the weighting mechanism for the central aggregation (RW-CA) process and local training (RW-LT) process. As shown in Table~\ref{table-ablation-all}, solely employing the RW-CA or RW-LT mechanism incurs an unstable performance gain. In Scenario 1, the RW-LT module marginally improves the Dice score but incurs a large performance drop in the second scenario. A similar phenomenon has been observed in~\cite{shen2021multi}, namely that re-weighting the training loss functions in each client generates unstable FL performance. For the RW-CA module, this introduces a slight performance drop based on the voxel-wise Dice score in the first scenario, while improving the segmentation accuracy under other metrics. Conversely, in the proposed FedMSRW framework, jointly incorporating the two re-weighting mechanisms consistently improves the FedBN method by a large margin, indicating the effectiveness and robustness of our method on the FL MS segmentation tasks.

\begin{table*}[!htb]
\centering
\caption{The experimental results on the brain tissue differences comparison (MSSEG daataset for the Scenario~1). }
\resizebox{0.7\linewidth}{!}{%
\begin{tabular}{c| c c c c | c c c c}
\hline
\multirow{2}{*}{} &  \multicolumn{4}{c|}{Grey Matter Difference (\%) $\downarrow$}  & \multicolumn{4}{c}{White Matter Difference (\%) $\downarrow$} \\
\cline{2-9}
 & $C1$ & $C2$  & $C3$ & $avg$ & $C1$  & $C2$ & $C3$ & $avg$   \\ \hline
 
 Single &
 $0.2418$ &  $0.3680$ &  $0.0417$  &  $0.2172$ & 
 $0.2640$ &  $0.4578$ &  $0.0323$  &  $0.2514$  \\
 
 Central &
 $0.2948$ &  $0.2338$ &  $0.0390$  &  $0.1892$ & 
 $0.3168$ &  $0.2938$ &  $0.0348$  &  $0.2151$  \\
 
 \hline
 FedAvg & 
 $0.7409$ &  $2.4815$ &  $\textbf{0.0419}$  &  $1.0881$ & 
 $0.8553$ &  $2.8010$ &  $\textbf{0.0212}$  &  $1.2258$  \\
 
 FedProx& 
 $0.7693$ &  $1.0077$ &  $0.0461$ &  $0.6077$ &  
 $0.8637$ &  $1.1711$ &  $0.0314$ &  $0.6887$ \\

 FedBN  & 
 $0.4303$ &  $0.2462$ &  $0.0482$ &  $0.2415$ &  
 $0.5190$ &  $0.3228$ &  $0.0494$ &  $0.3004$\\
 
 DWA  & 
 $0.4271$ &  $\textbf{0.1594}$ &  $0.0543$ &  $0.2256$ &  
 $0.4890$ &  $0.2787$ &  $0.0639$ &  $0.2772$ \\
 \hline
 
Ours & 
 $\textbf{0.3670} $ &  $0.1942$ &  $0.0550$ &  $\textbf{0.2054}$ &  
 $\textbf{0.4819}$ &  $\textbf{0.2785}$ &  $0.0542$ &  $\textbf{0.2715}$ \\
 \hline
\end{tabular}}
\label{table-brain-vol-s1}
\end{table*}

\begin{table*}[!htb]
\centering
\caption{The experimental results on the brain tissue differences comparison on Scenario~2. }
\resizebox{0.85\linewidth}{!}{%
\begin{tabular}{c| c c c c c| c c c c c}
\hline
\multirow{2}{*}{} &  \multicolumn{5}{c|}{Grey Matter Difference (\%) $\downarrow$}  & \multicolumn{5}{c}{White Matter Difference (\%) $\downarrow$} \\
\cline{2-11}
 & $C1$ & $C2$  & $C3$ & $C4$ & $avg$ & $C1$  & $C2$ & $C3$ & $C4$ & $avg$   \\ \hline
 
 Single &
 $0.0940$ &  $0.5978$ &  $0.3961$  &  $0.2400$ &  $0.2719$ & 
 $0.1281$ &  $0.8934$ &  $1.5420$  &  $0.2922$ &  $0.5978$   \\
 
 Central &
 $0.0722$ &  $0.0460$ &  $0.1762$  &  $0.3511$ &  $0.1552$ & 
 $0.1089$ &  $0.0791$ &  $0.2638$  &  $0.4287$ &  $0.2097$   \\
 
 \hline
 FedAvg & 
 $0.1999$ &  $0.0483$ &  $0.6747$  &  $0.4587$ &  $0.3393$ & 
 $0.3551$ &  $0.0656$ &  $0.8129$  &  $0.5921$ &  $0.4645$   \\
 
 FedProx  & 
 $0.2395$ &  $0.1707$ &  $0.1846$ &  $0.5318$ & $0.2815$ &  
 $0.7539$ &  $0.2884$ &  $0.6220$ &  $0.6605$ & $0.6314$\\

 FedBN  & 
 $0.1672$ &  $0.2160$ &  $0.1858$ &  $0.3293$ &  $0.2149$ &  
 $\textbf{0.2211}$ &  $0.4996$ &  $0.5437$ &  $0.4137$ &  $0.3789$\\
 
 DWA  & 
 $0.5005$ &  $1.1529$ &  $0.6983$ &  $0.4130$ & $0.6285$ &  
 $2.3295$ &  $2.1190$ &  $1.4647$ &  $0.5038$ & $1.6989$\\
 \hline
 
 Ours  & 
 $\textbf{0.1069}$ &  $\textbf{0.0325}$ &  $\textbf{0.1274}$ &  $\textbf{0.3188}$ &  $\textbf{0.1470}$ &  
 $0.2592$ &  $\textbf{0.0761}$ &  $\textbf{0.1696}$ &  $\textbf{0.3771}$ &  $\textbf{0.2370}$\\
 \hline
\end{tabular}}
\label{table-brain-vol-s2}
\end{table*}

\subsection{Different Model Design Strategies}

In this section, we present further experiments on the different design selections of the proposed re-weighting mechanism at the local and global levels. These experiments were conducted on both scenarios and the results are shown in Table~\ref{table-select-all}.

First, we replace the model's segmentation confidence in Equation~\ref{equ-rw-agg} with the entropy map of the whole segmentation predictions (`\textbf{Ours-ent}' in Table~\ref{table-select-all}), following typical uncertainty learning methods in medical image segmentation~\cite{liu2021adapting,yu2019uncertainty}. The Equation~\ref{equ-rw-agg} is then re-formulated as:

\begin{equation}
\begin{aligned}
P_{i}^{e} =  - M_{i}(x)* log \big(M_{i}(x)\big) * (1- L_{dice}\big(M_{i}(x), y\big)).
\label{equ-rw-agg-ent}
\end{aligned}
\end{equation}
 
Finally, each local model in the central aggregation process in Equation~\ref{equ-fedbn-rw-agg} is assigned a weight of $P_{i}^{e}$. Due to the severe imbalance of MS lesions in the brain MRI from the clinical practice, utilizing entropy maps incurs inaccurate representations of the model's segmentation confidence, and further degrades the FL segmentation performance in Scenario 2. Although the `\textbf{Ours-ent}' method achieves a slight performance gain in Scenario 1, we still select the global-level re-weighting mechanism based on the mask probability as defined in Equation~\ref{equ-fedbn-rw-agg}, due to the consistent performance gain.

In addition, we conducted experiments in which lesion volume was employed for local-level re-weighting on the task learning, referred to as the `\textbf{Ours-vol}' method in Table~\ref{table-select-all}. Specifically, the volume ratio $vr_{i}$ in Equation~\ref{equ-dice-rw} is replaced by the total number of lesion voxels $v_{i}$. Due to the inaccurate estimation of the true MS lesion distributions in brain MRI patches for model training, the `\textbf{Ours-vol}' method degrades the segmentation accuracies under voxel-level Dice in Scenario 1 and all metrics in Scenario 2. For both the `\textbf{Ours-ent}' and `\textbf{Ours-vol}' selections, we notice although they can improve the FedBN baseline in the Scenario~1, a severe performance drop has been incurred in the Scenario~2. We think the reasons for this phenomenon are two folds: 1) each client of the Scenario~2 has more data than those in Scenario~1; 2) the multi-client MS dataset in Scenario~2 is constructed by various datasets from in-house scanners and the public resources, which brings more distinctions for the cross-client data distributions.

\subsection{Evaluation on Brain Volumetric Analysis}

In addition to the lesion segmentation accuracy, we evaluated the impact of FL methods for lesion segmentation and inpainting on brain volumetric analysis, an important application in MS clinical trials and clinical practice. Essentialy, the presence of white matter MS lesions, which have an intensity approximating grey matter, leads to tissue misclassifications in brain volumetric analyses on T1 MR images. Lesion inpainting is therefore a routine pre-processing approach to remove the impact of lesions~\cite{tang2021lg} to brain tissue segmentation. However, the lesion inpainting methods are affected by the quality of the lesion segmentation masks. Thus, we compared brain volumetric analyses on T1 images inpainted using different segmentation approaches as an auxiliary analysis to further support the evaluation of the proposed method.

The pipeline in this section is based on our previous work~\cite{tang2021lg}. First, the intensity inhomogeneity in T1 weighted brain MR images of each subject was corrected using the N3 bias correction method from FreeSurfer~\cite{tustison2009n4itk}. Then FLAIR and T1 images were co-registered using FSL-FLIRT~\cite{jenkinson2002improved}. We applied the lesion inpainting method on T1 images separately using the corresponding MS lesion masks from ground truth and other compared methods, then FSL-FAST~\cite{zhang2001segmentation} was applied to segment the grey matter (GM) and white matter (WM) from the inpainted brain images.

The brain volumetric analysis results in both two scenarios are shown in Table~\ref{table-brain-vol-s1} and~\ref{table-brain-vol-s2}, where the grey matter and white matter percentage differences with respect to the whole brain tissue volume are included. Although other comparison methods achieve the best performance on WM or GM percentage difference on some specific clients, the proposed method outperformed all of them on the average performance on all clients in both scenarios. These auxiliary experiment results further support that the proposed method is capable of providing robust MS lesion masks that improve the performance of downstream image analysis tasks.

\section{Conclusion}

In this work, we propose a novel framework for FL MS lesion segmentation incorporating task-specific re-weighting mechanisms. Due to substantial inter-client variance in MS lesion data compared with typical FL settings, we observed limitations of previously described weighting mechanisms for central aggregation and local training. To this end, we first propose to re-weight the model aggregation process based on the segmentation ability for each local model, which alleviates the negative influence of local models with inferior segmentation ability on the central model. Considering the variance in lesion size distribution amongst clients, we further propose to re-weight the loss function in each local client based on the lesion volume ratio, avoiding model bias due to cross-client label distinctions. Extensive experiments in two FL MS lesion segmentation scenarios indicated the superiority of our proposed re-weighting mechanism compared with typical FL methods. In addition, brain volumetric analysis demonstrated the effectiveness of our proposed FL framework in practical research and clinical applications. The demand for privacy-preserving FL in clinical scenarios heightens the imperative to refine existing approaches. FedMSRW is an important methodological advance for analysing heterogenous multi-client imaging datasets with FL.

{\small
\bibliographystyle{IEEEtran}
\bibliography{IEEEabrv, ref}
}

\end{document}